\documentstyle[preprint,prc,aps,epsf]{revtex}
\begin{document}
\draft
\title{
Approximate treatment of electron Coulomb distortion in 
quasielastic $(e,e')$ reactions}
\author{ K.S. Kim,  L.E. Wright, and Yanhe Jin }
\address{
Institute of Nuclear and Particle Physics,
Department of Physics and Astronomy, Ohio University, Athens, 
Ohio 45701}
\author{ D.W. Kosik}
\address{Department of Physics, Butler University, Indianapolis, 
IN 46208 }

\maketitle
\begin{abstract}
In this paper we address the adequacy of various approximate methods 
of including Coulomb distortion effects in $(e,e')$ reactions by 
comparing to an exact treatment using Dirac-Coulomb distorted waves.
In particular, we examine approximate methods and analyses of 
$(e,e')$ reactions developed by Traini $et$ $al.$ using a high energy
approximation of the distorted waves and phase shifts due to Lenz and
Rosenfelder.
This approximation has been used in the separation of longitudinal 
and transverse structure functions in a number of $(e,e')$ 
experiments including the newly published $^{208}Pb(e,e')$ data from 
Saclay.
We find that the assumptions used by Traini and others are not valid 
for typical $(e,e')$ experiments on medium and heavy nuclei, and 
hence the extracted structure functions based on this formalism are 
not reliable.
We describe an improved approximation which is also based on the 
high energy approximation of Lenz and Rosenfelder and the analyses 
of Knoll and compare our results to the Saclay data.
At each step of our analyses we compare our approximate results to 
the exact distorted wave results and can therefore quantify the 
errors made by our approximations.
We find that for light nuclei, we can get an excellent treatment of 
Coulomb distortion effects on $(e,e')$ reactions just by using a 
good approximation to the distorted waves, but for medium and heavy 
nuclei simple additional $ad-hoc$ factors needs to be included.
We describe an explicit procedure for using our approximate analyses 
to extract so-called longitudinal and transverse structure functions 
from $(e,e')$ reactions in the quasielastic region.
\end{abstract}
\pacs{25.30.Fj 25.70.Bc}
\narrowtext

\section{Introduction}
Medium and high energy electron scattering has long been acknowledged 
as a useful tool in the investigation of nuclear structure and 
nuclear properties, especially in the quasielastic region.
In the plane wave Born approximation (PWBA), where electrons are 
described as Dirac plane waves, the cross section for inclusive 
quasielastic $(e,e')$ processes can be written simply as
\begin{equation}
\frac{d^2\sigma}{d\Omega_e d\omega}= \sigma_{M}
\{ \frac{q^4_\mu}{q^4}  S_L(q,w) + [ \tan^2 \frac{\theta_e}{2} -
\frac{q^2_\mu}{2q^2} ]  S_T(q,w) \} \label{pwsep}
\end{equation}
where $q_\mu ^2 = \omega^2-{\bf q}^2$ is the four-momentum transfer,
$\sigma _{M}$ is the Mott cross section given by $\sigma_{M} = 
(\frac{\alpha }{2E} )^2  \frac{\cos^2
\frac{\theta}{2}}{\sin^4 \frac{\theta}{2}}$, and $S_L$ and $S_T$ 
are the longitudinal and transverse structure functions which depend 
only on the momentum transfer $q$ and the energy transfer $\omega$.
By keeping the momentum and energy transfers fixed while varying the 
electron energy $E$ and scattering angle $\theta_e$, it is possible 
to extract the two structure functions with two measurements.
However, when the electron wavefunctions are not Dirac plane waves, 
but rather are distorted by the static Coulomb field of the target 
nucleus, such a simple formulation as given in Eq.~(\ref{pwsep}) is 
no longer possible and, in general, the cross section does not 
separate into the sum of longitudinal and transverse structure 
functions with coefficients which only depend on the electron 
kinematics. 
Clearly the size of the Coulomb distortion effects depends on the 
charge of the target nucleus and on the energy of the electrons. 
In general, Coulomb effects are not too large near the peaks of 
cross sections, but can have greatly magnified  effects when one 
extracts ``structure functions'' by subtracting one cross section 
from another.
For example, in a recent distorted wave calculation~\cite{jin94} 
of $^{16}O(e,e'p)$ in the quasielastic region,  we found Coulomb 
effects on the extracted spectroscopic factors to be approximately 
3\%, while the effects on the extracted fourth structure function 
was approximately 15\%.

However, some sort of extraction of structure functions, albeit 
approximate, is still very appealing since structure functions are 
sensitive to different aspects of the underlying knockout process or 
the final state interaction. Furthermore, $(e,e')$ reactions in the 
quasielastic region are particularly appealing in that the cross 
section is the sum of many knockout processes and some of the 
various Coulomb effects may be partially averaged out.
In fact, this is the case.
A rather simple approximation known as the Effective Momentum 
Approximation (EMA) where the electron momenta $p_{i,f}$ are 
modified by the value of the Coulomb potential at the center of the 
nucleus goes quite far in reproducing the Coulomb distortion effects 
for light nuclei.
However, the EMA for heavy nuclei does not adequately reproduce the 
DWBA cross section in the quasielastic region. 

During the past decade, $(e,e')$ cross sections have been 
measured~\cite{mez85}--~\cite{blat86} for a number of nuclei in the 
quasielastic region and either plane wave  or EMA was used to 
extract longitudinal and transverse structure functions.
These extracted structure functions were compared to the predictions 
of a simple Fermi-gas model, and in some cases, there appeared to be 
large suppression (up to about 40\%) of the longitudinal structure 
functions.
There were also disagreements with the extracted transverse 
structure functions and the predictions of the Fermi-gas model, but 
these were expected since exchange currents, pion production and 
other processes that are primarily induced by transverse photons 
were not included in the Fermi-gas model.
It should be noted that the Fermi-gas model is a rather crude 
description of a nucleus.
In particular, the shape of the structure functions at fixed 
momentum transfer plotted as a function of energy transfer is not 
well described.
On the other hand, we found that a ``single-particle'' relativistic 
model using relativistic Hartree wavefunctions coupled with the full 
DWBA treatment of Coulomb distortion for the electrons was in good 
agreement with the measured cross sections in the quasielastic 
region for $^{40}Ca$~\cite{jin92} and in reasonably good agreement 
for $^{238}U$.
Furthermore, the longitudinal structure function extracted for 
$^{40}Ca$ was well produced in magnitude and shape by this 
model~\cite{yates}.

More recently, Jourdan~\cite{jourdan} has examined the world data 
set for inclusive quasielastic scattering on $^{12}C$, $^{40}Ca$, 
and $^{56}Fe$ at momentum transfers of 300, 380 and 570 MeV/c, and 
has evaluated the Coulomb sum rule.
No evidence of suppression is found for the highest q value 
(570 MeV/c) where the sum rule is most model independent.
Jourdan used our Coulomb corrections in arriving at this conclusion.
Thus, on the basis of our direct comparison with the measured cross 
sections of $^{40}Ca$, and of Jourdan's results, the published 
analyses~\cite{saclay} for $^{208}Pb(e,e')$ which claim up to 50\% 
suppression of the longitudinal structure function is surprising.
However, the approximate treatment of Coulomb distortion used in the 
analysis is not accurate, and leads to doubts about the extracted 
structure functions.
As a separate matter, we question the  nuclear model used in making 
the claim of suppression.

In this paper we investigate the possibility of including Coulomb 
distortion effects in $(e,e')$ reactions in the quasielastic region 
for medium and heavy nuclei in an approximate way.
We have an advantage as compared to previous workers in that we have 
an exact treatment of the static Coulomb distortion of the target 
nucleus via a Distorted Wave Born Approximation (DWBA) calculation 
to which we can compare~\cite{jinphd}.
In Section II we will discuss various approximations that permit a 
``plane-wave-like'' approach to the treatment of Coulomb distortion 
and compare the approximate results to the exact DWBA results in a 
step by step way.
We obtain an approximate potential due to the electron current which 
describes Coulomb distortion quite well.
In Section III, we apply this potential with further approximations 
to the particular case of inclusive quasielastic processes.
Finally we compare our calculations to the Saclay data~\cite{saclay} 
for the inclusive reaction $^{208}Pb(e,e')$.

In addition, we give an explicit procedure for extracting 
longitudinal and transverse structure functions from inclusive cross 
section data in the quasielastic region from medium and heavy nuclei, 
and make some general conclusions.

\section{Approximations}
\subsection{Approximation of the Electron Potential}

The four potential arising from the electron current $j_{\mu}$ is 
simply given in terms of the retarded Green's function by
\begin{equation}
A_{\mu}({\bf r}) = \int  G({\bf r}_{e}, {\bf r}) j_{\mu}({\bf r}_{e}) 
d{\bf r}_{e}
\label{pot0}
\end{equation}
where
\begin{eqnarray}
G({\bf r}_{e}, {\bf r}) = {\frac {e^{{\imath} {\omega} {\mid {\bf r}
_{e} - {\bf r} \mid}}} {\mid {\bf r}_{e} - {\bf r} \mid}} \nonumber 
\end{eqnarray}
and the electron current is given in terms of the initial and final 
electron wavefunctions by
\begin{equation}
j_{\mu}({\bf r}_{e}) = \overline{\psi}_f({\bf r}_{e}) \gamma_\mu 
\psi_i({\bf r}_{e}).
\end{equation}
For scattering processes this current extends over all space and 
thus the integral in Eq.~(\ref{pot0}) is not straightforward.

Knoll proposed~\cite{knol} a way of replacing the integral in 
Eq.~(\ref{pot0}) by a series of differential operators by using the 
transformation,
\begin{equation}
\int V({\bf r} -{\bf r'})f({\bf r'}) d{\bf r'} = e^{{\imath}{\bf q'}
{\cdot}{\bf r}}
\tilde{V}(-{\bf q'} +\imath {\bf \nabla}) e^{-{\imath}{\bf q'}
{\cdot}{\bf r}} f({\bf r})
\end{equation}
between the function $V({\bf r}-{\bf r'})$ and its Fourier transform
$\tilde{V}({\bf q'})$.
Applying this result to Eq.~(\ref{pot0}) and making a Taylor series 
expansion, we obtain the following series expansion for the potential,
\begin{equation}
A_{\mu} ({\bf r}) = {\frac {4 {\pi}} {q'^{2}-{\omega}^{2}}} [1 + 
({\frac {q'^{2} + {\nabla}^{2}} {q'^{2} - {\omega}^{2}} }) + 
({\frac {q'^{2} + {\nabla}^{2}} {q'^{2} - {\omega}^{2}} })^{2} +
{\ldots} ] j_{\mu} ({\bf r}). \label{pot}
\end{equation}
Note that while the momentum variable $q'$ is arbitrary, the choice 
affects the convergence of the series.
In particular, for the case of Dirac plane waves for the electron 
the only dependence of the electron current on $\bf r$ is simply 
$e^{{\imath}{\bf q}{\cdot}{\bf r}}$ and choosing ${\bf q}'={\bf q}$ 
results in the vanishing of all the terms except the first, and 
Eq.~(\ref{pot}) reduces to the well known M${\ddot{o}}$ller potential,
\begin{eqnarray}
A_{\mu}({\bf r})={\frac {4{\pi}} {q^{2}-{\omega}^{2}} } {\bar{u}}
({\bf p}_{f}){\gamma}_{\mu}u({\bf p}_{i}) e^{{\imath}{\bf q}{\cdot}
{\bf r}}=a_{\mu}e^{{\imath}{\bf q}{\cdot}{\bf r}}, \nonumber
\end{eqnarray}
where $u$ is the familiar Dirac plane wave spinor.

As a test of this approximate procedure for calculating the potential,
 we calculated the electron charge distribution in the presence of 
the static Coulomb potential arising from the ground state charge 
distribution of $^{208}Pb$ using the partial wave solutions of the 
Dirac equation and evaluated the zeroth, zeroth plus first and zeroth
plus first and second terms in Eq.~(\ref{pot}).
Using these potentials we calculated the inelastic scattering cross 
section induced by a surface nuclear charge transition density
\begin{equation}
\rho^{if}_{n}({\bf r}) = {\frac {1} {R^{2}_{n}} } 
{\delta}(r-R_{n})Y^{M}_{L}({\hat r})
\end{equation}
where $R_{n}$ is the nuclear radius.
The ground state density of the nucleus was  described by a Fermi 
distribution of radius $R=6.65$ fm and total charge $Z=82$.
The cross sections were calculated at initial energy $E_{i}=400$ 
MeV and final energy $E_{f}=300$ MeV with energy transfer 
${\omega}=100$ MeV.
In agreement with Knoll~\cite{knol}, we found that the first and 
second correction terms are sufficient to fill up the minima and the 
contribution of the second correction term was less than $2 \%$ for 
momentum transfer $q {\geq}\;350$ MeV/c.
We conclude that this high momentum approximation provides an 
alternative procedure for calculating the potential arising from 
inelastic electron scattering processes.
However, this procedure does require the numerical solution of the 
Dirac equation using partial waves which does take some computational 
time.
One advantage, however, is that the radial functions only have to be 
calculated out to about three times the nuclear radius.

We also applied this procedure to various approximate solutions of 
the Dirac equation and noted that the first and second order  terms 
were not well controlled if one approximates the electron current.
Clearly one should not approximate a function and then differentiate 
it.
In the following when we examine approximate electron wavefunctions, 
and hence approximate electron currents, we will only use the zeroth 
term in Eq.~(\ref{pot}), but will be sensitive to the choice of $q'$.

\subsection{High Energy Wavefunction Approximation}

The incoming Coulomb distorted electron scattering wave function 
that satisfies appropriate boundary conditions for an electron with 
spin $s_i$ can be expressed in the form of a partial wave sum 
by~\cite{uber} 
\begin{equation} 
\Psi_{i}^{s_{i}}({\bf r}) = 4\pi{\sqrt {\frac{E_{i} + m}{2E_{i}}} }
\sum_{\kappa , \mu}e^{i\delta_\kappa} i^lC^{\ l\ \ 1/2\ \ j}_
{\mu-s_{i}\ s_{i}\ \mu}\ {Y_l^{\mu-s_{i}}}^*({\bf \hat p}_{i}) 
\psi^\mu_\kappa ({\bf r}),\label{iwfn}
\end{equation}
where the spinor $\psi_{\kappa}^{\mu}$ is an eigenstate with
angular momentum quantum numbers $\kappa$ and $\mu$ given explicitly 
by 
\begin{equation}
\psi_\kappa^\mu({\bf r}) =\left[\begin{array}{l} f_\kappa
(r)  \chi^\mu_\kappa({\bf \hat r}) \\ ig_\kappa(r)  
\chi^\mu_{-\kappa}
({\bf \hat r}) \end{array}\right], \label{wf1} 
\end{equation}
where the spin-angle functions are
\begin{equation}\chi^\mu_\kappa({\bf \hat r}) = \sum_{s} 
C^{\ l\ \ 1/2\ \ j}_{\mu-s\ s\ \mu}\ Y^{\mu-s}_l({\bf \hat r}) \chi_s. 
\label{wf2}
\end{equation}
The Dirac quantum number $\kappa$ determines the angular momentum 
labels for both $l$ and $j$.
Note that if we ignore the mass of the electron (valid for scattering 
angles away from extreme forward and backward angles),  the electron 
helicity is a good quantum number, and we only require positive 
$\kappa$ solutions.
The radial functions $f_\kappa$ and $g_\kappa$ and their
corresponding phase factors $\delta_\kappa$ are obtained by 
numerically solving the Dirac radial equation for a finite 
spherically symmetric nuclear charge distribution. 
The outgoing distorted wave function $\psi^{s_f}_f$ is found from 
Eq.~(\ref{iwfn}) by making the replacements $(i\delta_\kappa 
\rightarrow -i\delta_\kappa)$, $(s_i \rightarrow s_f)$,
$(E_{i} \rightarrow E_{f})$, and $({\bf p}_{i} \rightarrow 
{\bf p}_{f})$.
Note that by setting the phases $\delta_\kappa$ to zero and 
replacing $f_\kappa$ and $g_\kappa$ by the spherical Bessel 
functions $j_l(pr)$ and $sign(\kappa)j_{\bar l}(pr)$,  where $\bar 
l = l(-\kappa)$, the partial wave sum in Eq.~(\ref{wf1}) can be 
summed to give the Dirac plane wave solution,
\begin{equation}
\Psi_i({\bf r}) = {\sqrt {\frac {E_{i}+m} {2E_{i}}}} 
\left( \begin{array}{c}{\chi}_{s} \\ {\frac {{\mbox{\boldmath
$\sigma$}}{\cdot}{\bf p}} {E+m}}{\chi}_{s}
\end{array} \right)e^{i{\bf p}{\cdot}{\bf r}}, \label{plwv}
\end{equation}
where the electron spin label has been suppressed.  Thus, one way 
of obtaining a  plane-wave-like Coulomb distorted wavefunction, is 
to approximate the radial functions  $f_\kappa$ and $g_\kappa$  by 
spherical Bessel functions, and to approximate the scattering phase 
shifts  by an operator that can be pulled out of the partial wave 
sum.
Using these two ideas,  Lenz and Rosenfelder\cite{lenz} obtained an 
approximate scattering wavefunction for high energy electrons which  
includes Coulomb distortion for a finite nucleus in an approximate 
way more than twenty years ago.

In the high energy limit with good helicity $(E\gg m_{e})$, the 
approximate wavefunction of Lenz and Rosenfelder~\cite{lenz}, and 
Knoll~\cite{knol} can be written as
\begin{equation}
\Psi^{(\pm)} ({\bf r}) = {\eta} (r) e^{{\pm} {\imath} \delta 
({\bf J}^{2})} e^{{\imath} {\bf p} {\cdot} {\bf r} {\eta}(r)} u_{p} 
\label{hea}
\end{equation}
where $u_{p}$ is the Dirac plane wave spinor and
\begin{equation}
{\eta}(r)={\frac 1 {pr}}\int_{0}^{r} [p-V(r')]dr'
+\mbox{correction term}.
\end{equation}
Lenz and Rosenfelder approximated the dependence of the phases on 
the operator ${\bf J}^2$ by $\delta({\bf J}^{2}) = \delta_{1/2} + 
b({\bf J}^2 -3/4)$ where $b$ depends on the Coulomb potential.
This particular approximation for the phase shifts is not necessary 
since  any function of the operator ${\bf J}^2$ still allows the 
partial wave sum to be carried out.
Lenz and Rosenfelder claimed that this equation is valid for 
${\frac {{\mid}V(r){\mid}} {p}}{\ll}1$ and $j+1/2{\ll}pR$.
The second condition being primarily determined by the expression 
used for the phase shifts.

The approximate radial functions are given by spherical Bessel 
functions with modified argument,
\begin{eqnarray}
f_{\kappa}(r)&=&{\frac {x} {pr}}j_{l}(x) \nonumber \\
g_{\kappa}(r)&=&{\frac {x} {pr}}sign(\kappa)j_{{\bar l}}(x)
\end{eqnarray}
where $x=p^{\prime}(r)r$ and
\begin{equation}
p^{\prime}(r)=p-{\frac 1 r}{\int_{0}^{r}V(r)dr}+\frac{\Delta}{r} 
\label{loc}
\end{equation}
where $V(r)$ is the spherically symmetric Coulomb potential of the 
target nucleus and $\Delta$ is a correction term in the argument of 
order $\frac{\kappa^2}{(pr)^2}$ and is normally neglected.

To avoid having an $r$-dependent momentum, Traini $et$ 
$al.$~\cite{trai} made the further approximation that $p^{\prime}(r)
{\cong}p^{\prime}(0)=p-V(0)$ where $V(0)$ is the static Coulomb 
potential evaluated at the origin.
This approximation of the radial wavefunction leads to what is known 
as the Effective Momentum Approximation(EMA).
An approximation to the radial wavefunction where we keep the 
$r$-dependence in the momentum but still neglect $\Delta$  will be 
referred to as the Local Effective Momentum Approximation(LEMA).
In looking at these approximate wavefunctions, one should keep in 
mind that it is the spatial region around the nuclear surface that 
contributes most significantly to electron induced transitions.
That is, partial waves with angular momenta of order $pR$ where $R$ 
is the nuclear radius play a large role in the transition amplitude.
Thus, the approximation that the angular momentum is significantly 
less than $pr$ is not valid.  With this point in mind, we sought an 
$ad-hoc$ approximation for the correction $\Delta$.
We found that the following expression  describes the radial 
wavefunction at larger radial distances quite well,
\begin{eqnarray}
\Delta({\alpha}Z,E,{\kappa}^{2})=-{\alpha}Z({\frac {{16}{\kappa}} 
{E}})^{2}=a{\kappa}^{2}. \nonumber
\end{eqnarray}
The parameter $a=-{\alpha}Z({\frac {16} {E}})^{2}$ where the number 
16 is given in MeV and was determined by comparing to the exact 
distorted radial wavefunction.
We will label this case  LEMA + $\Delta$.  Note that we have chosen 
the parametrization of $\Delta$ so that it contains $\kappa^2$ 
which can be expressed in terms of the operator $J^2$ so that we 
will still be able to sum the partial wave series.

\begin{figure}[p]
\newbox\figa
\setbox\figa=\hbox{
\epsfysize=130mm
\epsfxsize=160mm
\epsffile{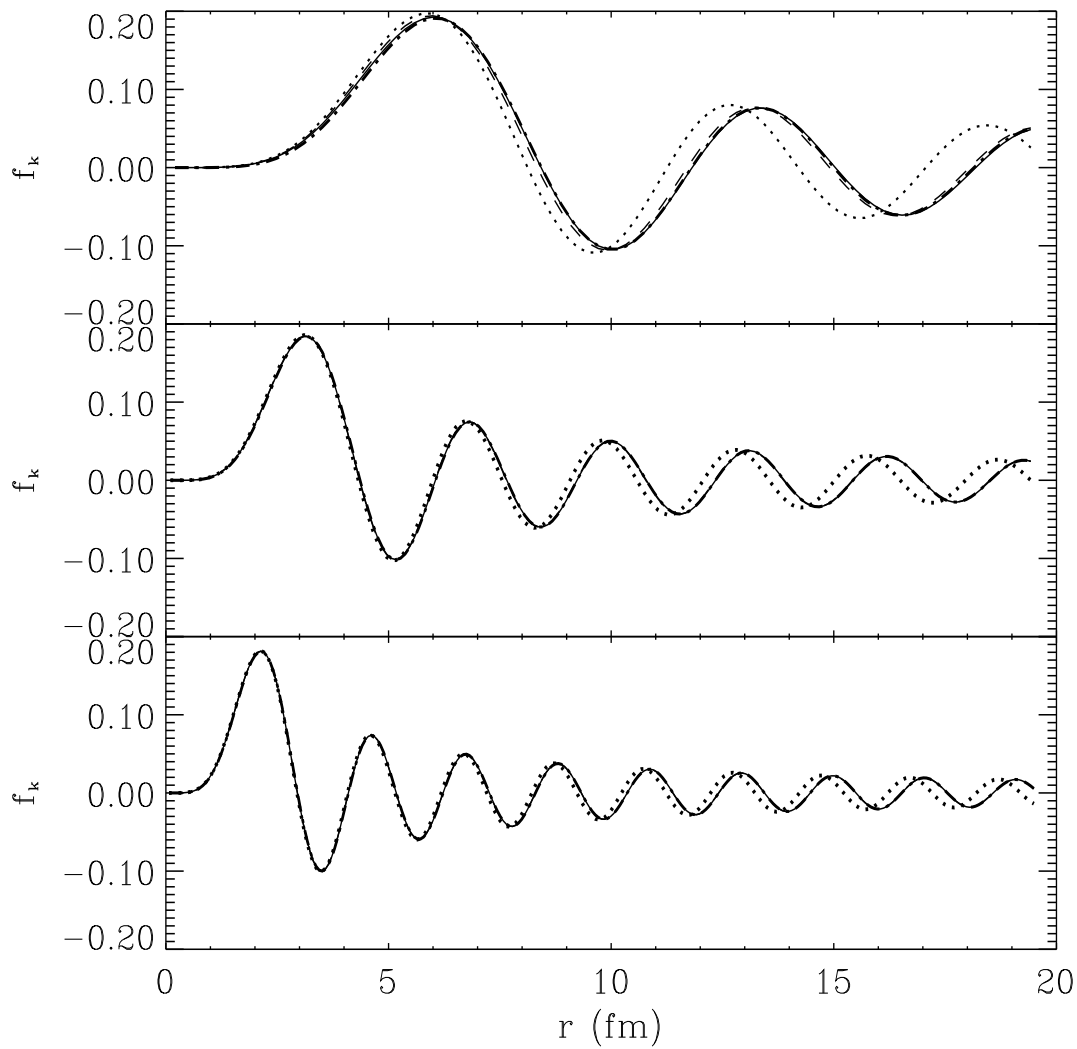}}
\noindent\hspace{0mm}\vspace{20mm}\box\figa
\caption{Radial wavefunctions in ${^{208}}Pb$ for ${\kappa}=5$. 
The top figure is for energy $E=200$ MeV, the middle for energy 
$E=400$ MeV and the bottom for energy $E=600$ MeV. The solid line 
is the exact DW wavefunction while the dash-dotted line is 
LEMA+$\Delta$, the dashed line is LEMA and the dotted line is EMA.}  
\label{radfun5}
\end{figure}

\begin{figure}[p]
\newbox\figb
\setbox\figb=\hbox{
\epsfysize=130mm
\epsfxsize=160mm
\epsffile{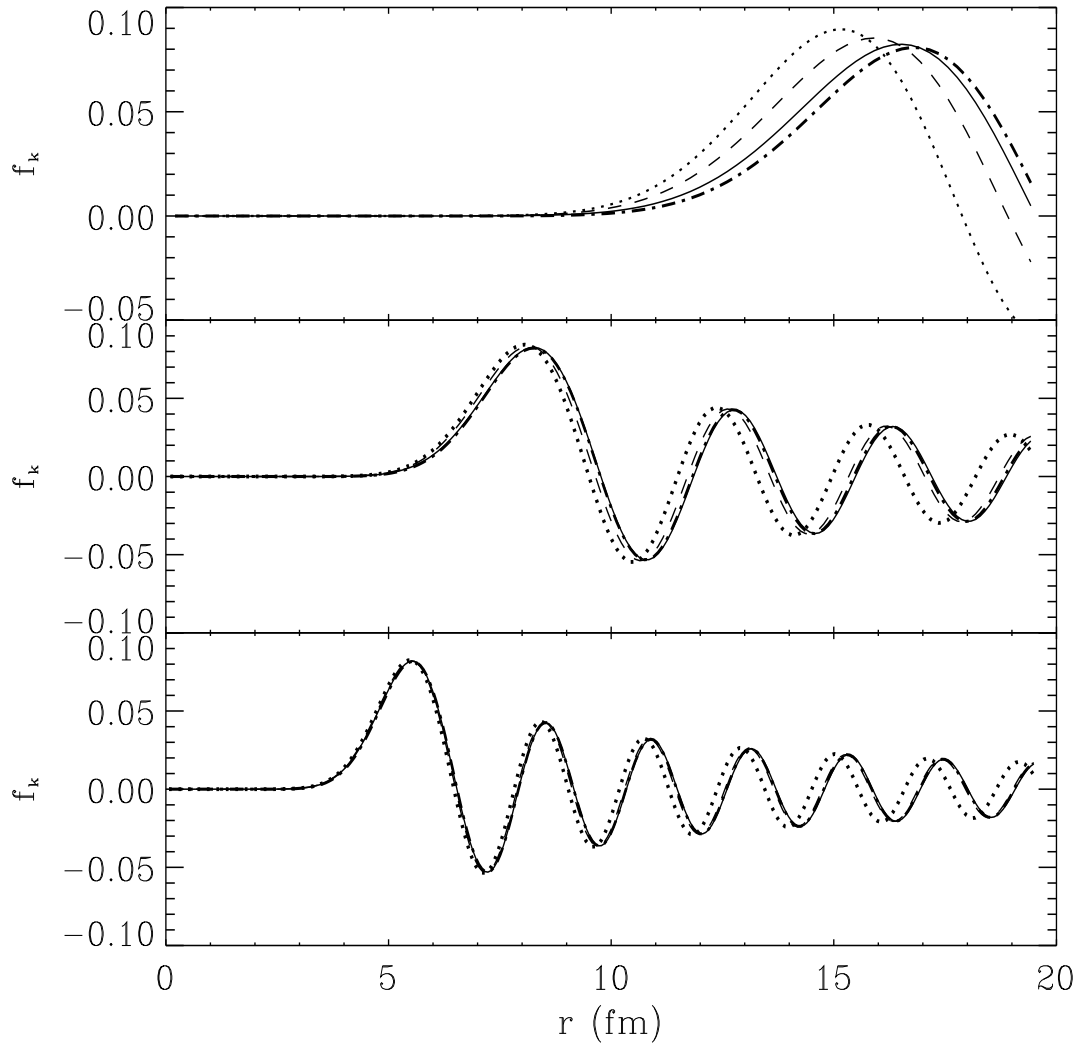}}
\noindent\hspace{0mm}\vspace{20mm}\box\figb
\caption{Radial wavefunctios in ${^{208}}Pb$ for ${\kappa}=15$. 
The same as Fig.~(\ref{radfun5}).}
\label{radfun15}
\end{figure}

To investigate these approximations we compare the radial 
wavefunctions calculated using EMA, LEMA, and LEMA + $\Delta$ to 
the exact Coulomb distorted waves (DW) for various angular momentum 
states for electron scattering from $^{208}Pb$.
The radius of $^{208}Pb$ is approximately 6.5 fm.
Figs.~(\ref{radfun5}) and ~(\ref{radfun15}) show the comparison 
between three approximate radial wavefunctions and the exact 
distorted radial wavefunction for different energies ($E=200$ MeV, 
$400$ MeV and $600$ MeV) and different $\kappa$ values($\kappa=5$ 
and $\kappa=15$).
The EMA wavefunction is acceptable at small radial distances, 
particularly much less than the nuclear radius, but at larger radii 
the approximate radial wavefunction is shifted too much to the left 
indicating the potential $V(0)$ at the origin is too large.
The LEMA wavefunction is a much better approximation to the exact 
wavefunction than EMA for medium and high energies out to radii 
beyond the nuclear radius, but at lower energies ( less than 250 
MeV), it also deviates significantly from the DW result.  
The wavefunction with the LEMA +$\Delta$ correction agrees with the 
distorted one almost perfectly above  400 MeV.

At the lower energy this wavefunction is a bit to the right  of the 
distorted wavefunction, but the discrepancy is acceptably small.
Our conclusion is that LEMA is much better than EMA and may be 
acceptable for some reactions. The approximation LEMA + $\Delta$ 
furnishes quite a good description of the distorted wave radial 
functions out to several nuclear radii.

\begin{figure}[p]
\newbox\figc
\setbox\figc=\hbox{
\epsfysize=130mm
\epsfxsize=160mm
\epsffile{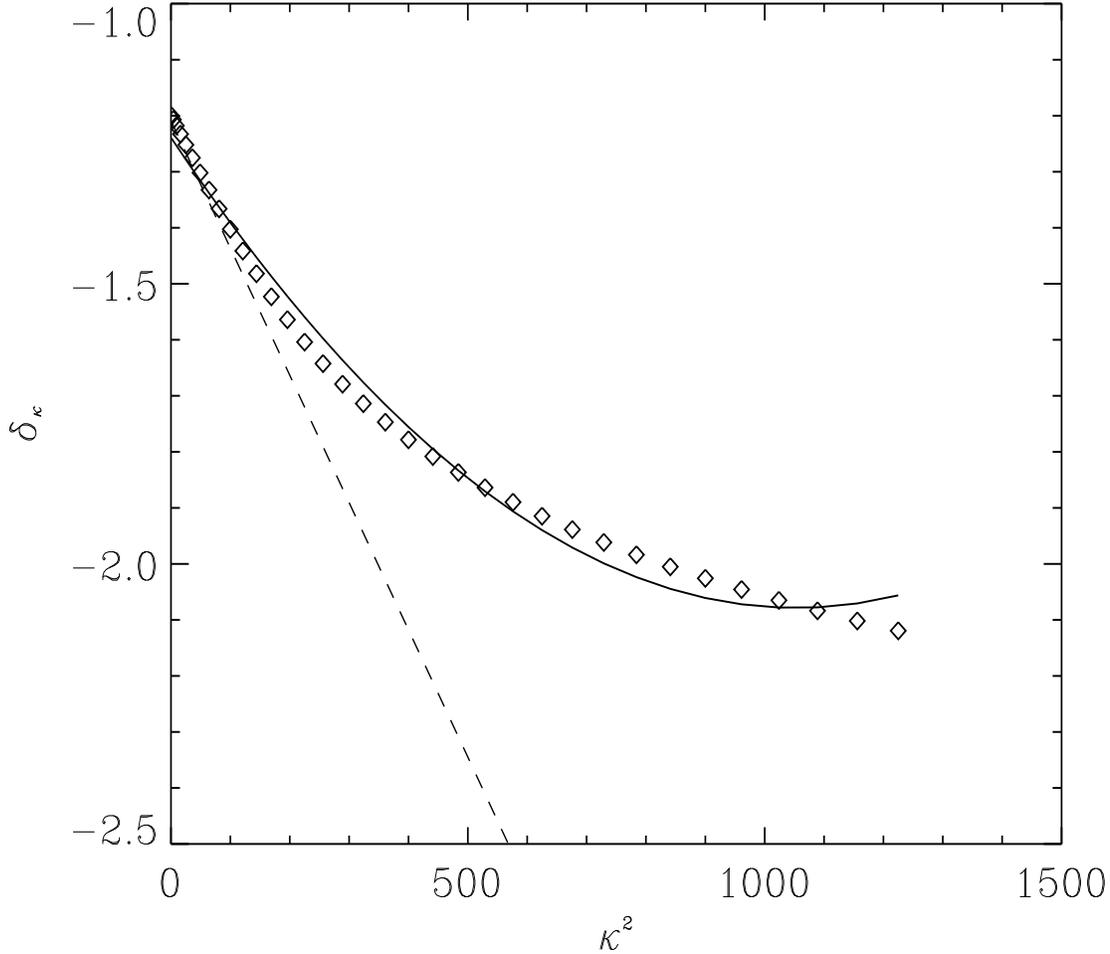}}
\noindent\hspace{-5mm}\vspace{30mm}\box\figc
\caption{Comparison of the exact and ${\delta}({\kappa}^{2})$ fits 
to the phases in ${^{208}}Pb$ for the energy $E=400$ MeV and 
${\kappa}_{max}=35$. 
The diamonds are the exact phases and the dashed line is the 
${\kappa}^{2}$-fit to the phases and the solid line includes the 
${\kappa}^{4}$ term in the fit.}
\label{phase}
\end{figure}

The other important ingredient in Coulomb distortion considerations 
are the phase shifts ${\delta}_{\kappa}$.
In order to sum the partial wave series in Eq.~(\ref{iwfn}), the 
phase shifts in  Eq.~(\ref{hea}) were expressed as a function of the 
total angular momentum operator $J^{2}$.
The eigenvalues for this operator are $j(j+1)$ when operating on the 
partial waves, so the question is how well are the exact phase 
shifts reproduced by the expression~\cite{knol},
\begin{eqnarray}
{\delta}_{j}&=& {\delta}_{1/2}+b[j(j+1)-{\frac 3 4}]\nonumber \\
&=&{\delta}_{1/2}+b[{\kappa}^{2}-1] \label{phs1}
\end{eqnarray}
where we used the relation $j = \mid\kappa\mid-1/2$.
For a uniform charge distribution of radius R,
\begin{equation}
{\delta}_{{\frac 1 2}}=Z{\alpha}({\frac 4 3}-{\ln {2pR}})+b 
\end{equation}
and
\begin{equation}
b=-{\frac {3Z{\alpha}} {4{p(0)^{\prime}}^{2}R^{2}}}.
\end{equation}
We investigated this ${\kappa}^{2}$ approximation for the phase 
shifts for the Coulomb potential with ${^{208}}Pb$ and 400 MeV 
electrons.
The phase shifts are in good agreement with the exact phase shifts 
for small $\kappa$ values, but for large $\kappa$ values, the 
approximate phase shifts are much too large in magnitude.
The breakdown of the approximation occurs for ${\kappa}{\approx}pR$ 
as expected.
However, as noted earlier, it is these orbitals that play a dominant 
role in electron induced reactions from the nucleus.
In order to avoid this violation, we assume that the approximate 
phase shifts contain a ${\kappa}^{4}$-term expressed as
\begin{eqnarray}
{\delta}_{\kappa}&=&a_{0}+a_{2}{\kappa}^{2}+a_{4}{\kappa}^{4} 
\nonumber \\
&=&b_{0}+b_{2}[j(j+1)-{\frac 3 4}]+b_{4}[j(j+1)-{\frac 3 4}]^{2} 
\label{phs2}
\end{eqnarray}
where $b_{0}=a_{0}+a_{2}+a_{4}$, $b_{2}=a_{2}+2a_{4}$, and 
$b_{4}=a_{4}$.
Unfortunately, we do not have a simple analytical expression for 
the coefficients, so $a_{0}$, $a_{2}$ and $a_{4}$ are  fitted to the 
exact phase shifts calculated with a distorted wave code.
As is evident in Fig.~(\ref{phase}), including the ${\kappa}^{4}$ 
terms leads to a much better description of the exact phases.

Clearly higher powers of $\kappa^2$ could also be included if needed.
Of course, using the description of the phases given in 
Eq.~(\ref{phs2}) rather than the one in Eq.~(\ref{phs1}) requires 
calculation of the exact Coulomb phases for both the incoming and 
outgoing electron energy and then determining the coefficients $a_n$ 
by a fitting procedure.
However, the solution for the phases is straightforward and very 
rapid for modern computers, so this poses no real practical problem.

We have calculated various multipoles of the scalar potential in the 
partial wave formalism and confirmed that the approximate radial 
function and the approximate phases are in close agreement with the 
exact potential.
These two approximations permit the summation of the partial wave 
series of Eq.~(\ref{iwfn}) so that the electron wavefunction for 
incoming or outgoing waves can be written as
\begin{equation}
\Psi^{({\pm})}({\bf r})={\frac {p'(r)} {p}}e^{{\pm}i{\delta}({\bf
J}^{2})}e^{i(1+{\frac {\Delta} {p'(r)r}}){\bf p}'(r){\cdot}
{\bf r}}u_{p}.
\label{w-k1}
\end{equation}
The operator ${\bf J}^{2}$ is the square of the total angular 
momentum operator $({\bf J}={\bf L}+{\bf S})$, and previous 
workers~\cite{trai,giupac} have made a power series expansion of 
the phase term and applied successive terms to the plane-wave-like 
part of the wavefunction.
The problem with this procedure is that the phase shifts are greater 
than one for $\kappa$ values of importance and this series is very 
slowly converging. 
It is straightforward to check that keeping only the first three 
terms in the exponential expansion leads to significant errors. 
We choose not to make this expansion, but to approximate $J^2$ by 
the orbital angular momentum squared $L^2$ and further to replace 
$L^2$ by its classical value 
$({\bf r}{\times}{\bf p}(r)^{\prime})^{2}$. 
Clearly we are neglecting the spin dependence of the phase shifts 
by this classical approximation, but since the processes we are 
interested in are dominated by angular momentum values around 10 or more, 
we expect the spin dependence to be negligible. 
We confirmed this estimate by comparisons of our partial wave calculation
of $(e,e'p)$ where the full spin dependence is included to a three
dimensional numerical integration using the classical approximation.
In like manner, we also replace the $\kappa^2$ dependence in 
$\Delta$ by $({\bf r}{\times}{\bf p}(r)^{\prime})^{2}$.
Finally, the approximate Coulomb distorted wavefunction of 
Eq.~(\ref{w-k1}) with the $\Delta$ correction is given explicitly 
by~\cite{kimphd}
\begin{equation}
\Psi^{({\pm})}({\bf r})={\frac {p^{\prime}(r)} {p}}e^{{\pm}i{\delta}
(({\bf r}{\times}{\bf p}^{\prime}(r))^{2})}e^{ia({\hat p}^{\prime}
(r){\cdot}{\hat r})({\bf r}{\times}{\bf p}^{\prime}(r))^{2}}
e^{i{\bf p}^{\prime}(r){\cdot}{\bf r}}u_{p}.
\label{w-k2}
\end{equation}
with ${\delta}(({\bf r}{\times}{\bf p}'(r))^{2})=b_{0}+b_{2}
({\bf r}{\times}{\bf p}'(r))^{2}+b_{4}({\bf r}{\times}
{\bf p}'(r))^{4}$.
Using this wavefunction and the first term of Eq.~(\ref{pot}) we 
obtain the following four potential, which includes in an 
approximate way, the Coulomb distortion of the target nucleus
\begin{equation}
A^{\mu}({\bf r})={\frac {4{\pi}} {4p_{i}p_{f}{\sin^4 {\frac 
{{\theta}_{e}} {2}}}}}e^{i[{\delta}_{i}(({\bf r}{\times}{\bf p}'
_{i}(r))^{2})+{\delta}_{f}(({\bf r}{\times}{\bf p}'_{f}(r))^{2})]}
e^{i({\Delta}_{i}-{\Delta}_{f})}e^{i{\bf q}'(r){\cdot}{\bf r}}
{\bar u}_{f}{\gamma}^{\mu}u_{i} \label{apppot}
\end{equation}
where $\Delta=a({\bf r}{\times}{\bf p}'(r))^{2}({\hat r}{\cdot}
{\hat p}'(r))$ and ${\bf q}'(r)={{\bf p}'}_{i}(r)-{{\bf p}'}_{f}(r)$.
The approximate potential of Eq.~(\ref{apppot}) is similar to the 
plane wave result except for the phase factors and the radial 
dependence in the momentum transfer.
Unfortunately, the spatial dependence in the phase factors makes a 
multipole decomposition of this potential impractical.
However, since it is an analytical function it is straightforward 
to calculate interaction matrix elements by performing the three 
dimensional integration over ${\bf r}$ numerically.

\section{Application to the Inclusive Process}

For the inclusive cross section $(e,e')$, the longitudinal and 
transverse structure functions in Eq.~(\ref{pwsep}) are bi-linear 
products of the Fourier transform of the components of the nuclear 
transition current density integrated over outgoing nucleon angles.
Furthermore, it is the Dirac structure of the M\"{o}ller 
potential which leads to the cross section containing one term with 
only longitudinal components of the current and a second containing 
only transverse components.
However, the Dirac structure of the approximate potential in 
Eq.~(\ref{apppot}) is the same as the plane wave result.
Therefore, even with Coulomb distribution included, albeit in an 
approximate way, the separation of the cross section into a 
longitudinal term and a transverse term persists. 
Explicitly, the structure functions for knocking out nucleons from a 
shell with angular momentum $j_{b}$ are given by
\begin{eqnarray}
S_{L}(q,{\omega})&=&\sum_{{\mu}_{b}s_{P}}{\frac {{\rho}_{P}} 
{2(2j_{b}+1)}} \int {\mid}N_{0}{\mid}^{2}d{\Omega}_{P} \\
S_{T}(q,{\omega})&=&\sum_{{\mu}_{b}s_{P}}{\frac {{\rho}_{P}} 
{2(2j_{b}+1)}} \int
({\mid}N_{x}{\mid}^{2}+{\mid}N_{y}{\mid}^{2})d{\Omega}_{P}
\end{eqnarray}
where the nucleon density of states ${\rho}_{P}={\frac {pE_{p}} 
{(2\pi)^{2}}}$, the $z$-axis is taken to be along ${\bf q}$, and 
${\mu}_{b}$ and $s_{P}$ are the z-components of the angular momentum 
of the bound and continuum state particles.
The Fourier transfer of the nuclear current $J^{\mu}({\bf r})$ is 
simply,
\begin{equation}
N^{\mu}=\int J^{\mu}({\bf r})e^{{\imath}{\bf q}{\cdot}{\bf r}}d^{3}r.
\end{equation}
The continuity equation has been used to eliminate the $z$-component 
($N_{z}$) via the equation $N_{z}=-{\frac {\omega} {q}}N_{0}$.
Note that when we use the approximate electron four potential along 
with current conservation to eliminate the z-component of the current 
we run into a problem since the momentum transfer ${\bf q}^{\prime}$ 
depends on r both in magnitude and direction.
In addition, the phase factors depend on ${\bf r}$.
To avoid generating additional terms we assume the direction of 
${\bf q}^{\prime}(r)$ is along the asymptotic momentum transfer 
${\bf q}$ which defines the ${\hat z}$-axis, and neglect the 
dependence on ${\bf r}$ in the phases and in ${\bf q}^{\prime}(r)$, 
when taking the divergence of ${\bf N}$.
With this further approximation, current conservation implies 
${\omega}N_{0}+{\bf q}^{\prime}(r){\cdot}{\bf N}=0$.
Using the approximate potential of Eq.~(\ref{apppot}), the cross 
section with for the inclusive reaction $(e,e')$ can be written as
\begin{equation}
\frac{d^2\sigma}{d\Omega_e d\omega}= \sigma_{M}
\{ \frac{q^4_\mu}{q^4}  S_L(q',w) + [ \tan^2 \frac{\theta_e}{2} -
\frac{q^2_\mu}{2q^2} ]  S_T(q',w) \} 
\end{equation}
and the transform of the transition nuclear current elements which 
appears in $S_{L}$ and $S_{T}$ are given by
\begin{eqnarray}
N_{0}&=&\int ({\frac {q'_{\mu}(r)} {q_{\mu}}})^{2} 
({\frac {q} {q'(r)}})^{2}e^{i{\delta}_{f}([{\bf r}{\times}
{\bf p}_{i}^{\prime}(r)]^{2})}e^{i{\delta}_{f}([{\bf r}{\times}
{\bf p}_{f}^{\prime}(r)]^{2})}e^{i({\Delta}_{i}-{\Delta}_{f})}
e^{i{\bf q}^{\prime}(r){\cdot}{\bf r}}J_{0}({\bf r})d^{3}r 
\label{apn0}\\
{\bf N}_{T}&=&\int e^{i{\delta}_{i}([{\bf r}{\times}
{\bf p}_{i}^{\prime}(r)]^{2})}e^{i{\delta}_{f}([{\bf r}{\times}
{\bf p}_{f}^{\prime}(r)]^{2})}e^{i({\Delta}_{i}-{\Delta}_{f})}
e^{{\imath}{\bf q}^{\prime}(r){\cdot}{\bf r}}J_{T}({\bf r})d^{3}r 
\label{apnt}
\end{eqnarray}

As noted earlier, a multipole expansion of the approximate potential 
is not practical, and since the inclusive reaction $(e,e')$ requires 
a sum over all occupied neutron and proton shells, numerical 
integration is very time consuming.
In order to have a more practical procedure we choose to make 
additional approximations.
Firstly, we neglect all of the phase factors($\delta$ and $\Delta$) 
in Eqs.~(\ref{apn0}) and ~(\ref{apnt}) but retain the $r$-dependence 
in ${\bf q}^{\prime}$.
This returns us to the approximation we call LEMA.
For light to medium nuclei this approximation is in good agreement 
with the full DWBA result regarding the shape as a function of 
energy transfer, but has a small discrepancy in magnitude. 
The magnitude is corrected with an overall factor of 
$({\frac {p_{i}'(0)} {p_{i}}})^{2}$ in the cross section.
However, for heavier nuclei, we noticed that LEMA with the magnitude 
factor was a very good 
approximation to the DWBA results for large electron scattering 
angle where the transverse term dominates, but deviated significantly 
for forward electron angles where the longitudinal term has a 
significant contribution.
Thus it appears that the phase factors play a significant role in 
the longitudinal term for large Coulomb distortion.

We also noticed that for forward electron angles the low energy 
side of the DWBA quasielastic peak looks similar to the plane wave 
result.
Thus, it appears that in the longitudinal term the phase factors are 
partially cancelling the effect of the effective momentum 
${\bf q}^{\prime}(r)$ on the low $\omega$ side of the quasielastic
 peak.
We examined a number of simple $ad-hoc$ modifications to the longitudinal
term as described by LEMA and 
the magnitude factor in 
order to reproduce the DWBA forward angle results for $(e,e')$ 
reactions on $^{208}Pb$ in the quasielastic region.
Based on a number of trials we propose the following ``Fourier'' transform for the charge 
component of the current 
\begin{equation}
N^{LEMA^{\prime}}_{0}=({\frac {p_{i}'(0)} {p_{i}}})
\int e^{i{\bf q}''(r){\cdot}{\bf r}}J_{0}
({\bf r})d^{3}r \label{lemal'}
\end{equation}
where $({\frac {p_{i}'(0)} {p_{i}}})$ is the magnitude 
enhancement and ${\bf q}''(r)={\bf p}_{i}''(r)-{\bf p}_{f}''(r)$, 
$p''(r)=p-{\frac {\lambda} {r}}\int_{0}^{r}V(r')dr'$ and the 
factor $\lambda$, which depends on the energy transfer $\omega$, 
is given by $\lambda=({\omega}/{\omega}_{0})^{2}$ with 
${\omega}_{0}={\frac {q^{2}} {1.4M}}$.
Clearly for small $\omega$, ${\bf q}''$ approaches the asymptotic 
value, while for $\omega$ past the quasielastic peak which is located 
approximately at ${\omega}_{0}$ in the cross section, the effective 
momentum differs significantly from the asymptotic value.
We have tested this $ad-hoc$ prescription for 
$0.3{\leq}{\omega}/{\omega}_{0}{\leq}2.0$ for a range of energies 
and nuclei and find excellent agreement with the DWBA result.
The transverse ``Fourier'' transform only contains the normalization  
factor and is given by
\begin{equation}
{\bf N}^{LEMA^{\prime}}_{T}=({\frac {p_{i}'(0)} {p_{i}}})
\int e^{i{\bf q}'(r){\cdot}{\bf r}}J_{T}({\bf r})
d^{3}r. \label{lemat'}
\end{equation}
We will refer to the cross section for the inclusive reaction 
calculated with these two structure functions as the 
LEMA$^{\prime}$ result.
Clearly $N^{LEMA^{\prime}}_{0}$ and $N^{LEMA^{\prime}}_{T}$ 
represent a modified Fourier transform 
of the nuclear transition current.
The approximation known as the EMA replaces ${\bf q}'(r)$ with 
${\bf q}'(0)$ wherever it appears in Eqs.~(\ref{apn0}) and 
~(\ref{apnt}) for $N_{0}$ and $N_{T}$ and the phases are neglected 
as usual.
We find that for light nuclei the EMA is adequate, but it leads to 
large errors for nuclei as heavy as $^{208}Pb$.

In our analyses of quasielastic scattering, we use relativistic 
bound and continuum single particle wavefunctions.
For the inclusive reaction we use continuum solutions for the 
outgoing, but unobserved, nucleon which are in the same Hartree 
potential as the bound state orbitals~\cite{jin92}.
This choice insures charge conservation and gauge invariance.
Thus, in this relativistic single particle model the nuclear 
current matrix element is
\begin{equation}
J^{\mu}({\bf r})=e{\bar {\Psi}}_{P}{\hat J}^{\mu}{\Psi}_{b}
\end{equation}
where we use the free nucleon current operator 
\begin{equation}
{\hat J}^{\mu}=F_{1}{\gamma}^{\mu}+F_{2}{\frac {i{\mu}_{T}} 
{2m_{N}}}{\sigma}^{{\mu}{\nu}}q_{\nu}
\end{equation}
where $\mu_{T}$ is the nucleon anomalous magnetic moment 
(for proton $\mu_{T}=1.793$ and for neutron $\mu_{T}=-1.91$).
The form factor $F_{1}$ and $F_{2}$ are related to the electric 
and magnetic form factors $G_{E}$ and $G_{M}$ by
\begin{eqnarray}
G_{E}&=&F_{1}+{\frac {{\mu}_{T}q_{\mu}^{2}} {4M^{2}}}F_{2} \\
G_{M}&=&F_{1}+{\mu}_{T}F_{2}
\end{eqnarray}
We choose the standard result:
\begin{equation}
G_{E}=G_{M}/({\mu}_{T}+1)=(1-q^{2}_{\mu}/0.71)^{-2}
\end{equation}
where in this formula $q_{\mu}^{2}$ is in units of GeV$^{2}$.

\begin{figure}[p]
\newbox\figda
\setbox\figda=\hbox{
\epsfysize=120mm
\epsfxsize=155mm
\epsffile{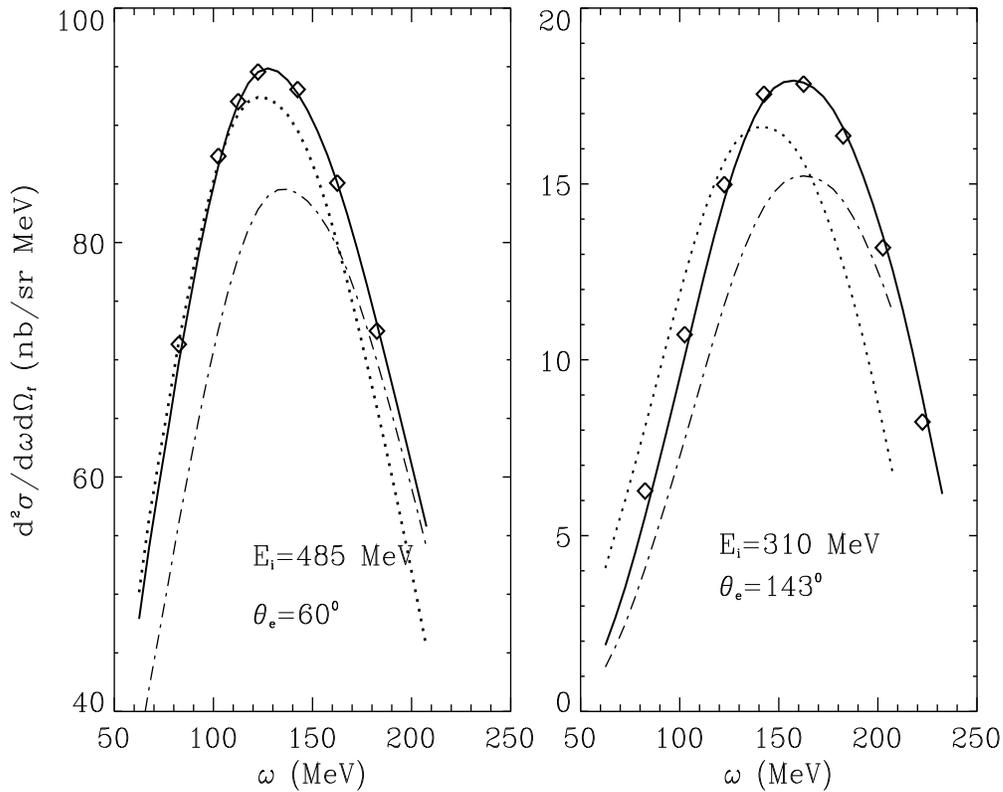}}
\noindent\hspace{5mm}\vspace{10mm}\box\figda
\caption{The differential cross section for ${^{208}}Pb(e,e')$ 
at two different electron energies and scattering angles. 
The dotted line is the PWBA result, the diamonds are the DWBA result, 
the dash-dotted line is the EMA result, and the solid line is the 
LEMA$^{\prime}$ result.}
\label{cros}
\end{figure}

In Fig.~(\ref{cros}), we compare various approximations to the DWBA 
result as a function of the energy transfer $\omega$ for two cases, 
incident electron energy $E_{i}=310$ MeV, scattering angle 
$\theta_{e}=143^{o}$ and $E_{i}=485$ MeV, scattering angle 
$\theta_{e}=60^{o}$ data sets.
The dotted line is the PWBA result, the diamonds are the DWBA result, 
the dash-dotted line is the EMA result, and the solid line is the 
LEMA$^{\prime}$ result.
We notice that the EMA result is always lower than the DWBA result 
although the peak position is approximately in the right place.
This lack of agreement with the EMA is not too surprising based on 
our previous examination of the wavefunction.
Replacing the average value of the Coulomb potential between the 
origin and the position $r$ by the value at the origin is too large 
an error for $r$ near the nuclear surface where most of the 
interaction takes place.
Since the EMA is such a bad approximation to the full DWBA result, 
it should not be used as a basis for including the Coulomb phase terms in 
$(e,e')$ or $(e,e'p)$ reactions.
Over the whole region, the LEMA$^{\prime}$ result is in excellent 
agreement with the DWBA result apart from the extreme wings of the 
quasielastic peak.
The difference around the peak is less than 2 $\%$ and side parts are 
about 5 $\%$.

\begin{figure}[p] 
\newbox\figga
\setbox\figga=\hbox{
\epsfysize=120mm
\epsfxsize=155mm
\epsffile{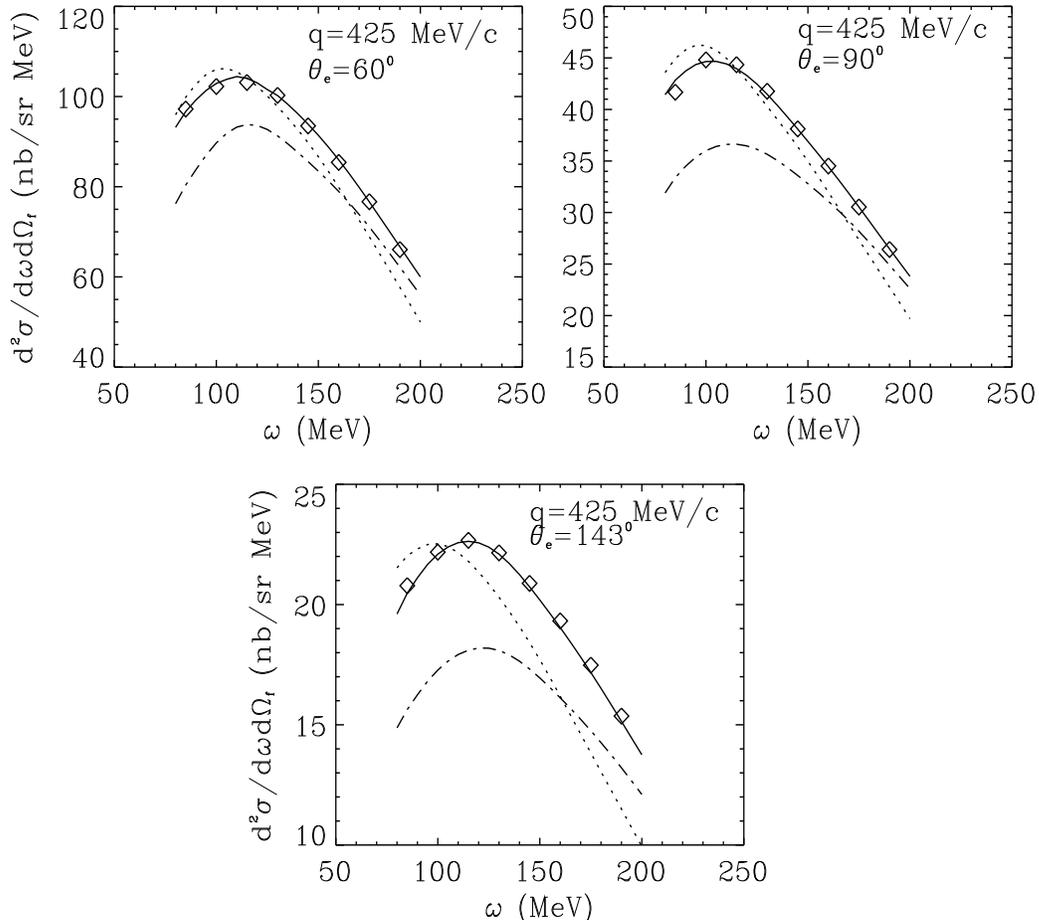}}
\noindent\hspace{5mm}\vspace{5mm}\box\figga
\caption{The differential cro10s section for $^{208}Pb$ at constant
momentum transfer $q=425$ MeV/c, but three different electron 
scattering angles.}
\label{croqfx}
\end{figure}

We illustrate our approximations for other kinematics by calculating 
the cross section at fixed momentum transfer $q=425$ MeV/c with 
three different electron scattering angles (${\theta}_{e}=60^{0}$,
$90^{0}$, and $143^{0}$) as shown Fig.~(\ref{croqfx}). 
The EMA result is always lower than the DWBA and the LEMA$^{\prime}$ 
results and is shifted toward large energy transfer by about 10 MeV. 
LEMA$^{\prime}$ again reproduces the DWBA cross sections quite well.

\section{Separation Procedure and Structure Functions}

Since the LEMA$^{\prime}$ cross section for $(e,e')$ has the same 
structure as the plane wave result, a Rosenbluth-type separation 
can be used to extract a longitudinal and a transverse contribution 
even in the presence of large Coulomb distortions.
In fact,  experimental evidence for this was given in one of the 
early papers on
quasielastic scattering~\cite{blat86} where cross section 
measurements on $^{238}U$ at three different electron scattering 
angles, but with the same energy and momentum transfer, fell on a 
straight line when a Rosenbluth plot was made.
Our LEMA$^{\prime}$ approximation gives a theoretical explanation 
for this observation.
Of course the separated structure functions are no longer bi-linear 
products of the simple Fourier transforms of the current components 
integrated over outgoing nucleon directions.
Inclusion of Coulomb distortion within LEMA$^{\prime}$ results in an 
$r$-dependent Fourier momentum variable which differs for the 
longitudinal and transverse case.
We have only been able check LEMA$^{\prime}$ for our particular model 
of the quasielastic process, so we can not prove that it applies to 
other models.  
However, based on previous work we know our model describes $(e,e'p)$ 
and $(e,e')$ quite well in the quasielastic region and its treatment 
of the spatial dependence in the nuclear charge and current is 
very realistic.
Thus we believe that the Coulomb corrections that we calculate 
will be appropriate for any realistic nuclear model.
We conclude that LEMA$^{\prime}$ is a good approximation for 
the inclusive cross section $(e,e')$ in the quasielastic region.

To illustrate the ``Rosenbluth'' separation we write 
$S = S_L + x S_T$ where S is the total structure function given by
\begin{equation}
S=({\frac {q} {q_{\mu}}})^{4} {\frac 1 {{\sigma}_{M}}}
{\frac {d^{2}{\sigma}} {d{\Omega}_{e}d{\omega}}}
\end{equation}
and $x=[ {\tan^2} {\frac{{\theta}_{e}} {2}} - {\frac {q^{2}_{\mu}} 
{2q^{2}}} ]/{\frac {q^{4}_{\mu}} {q^{4}}}$.

\begin{figure}[p] 
\newbox\figxa
\setbox\figxa=\hbox{
\epsfysize=200mm
\epsfxsize=155mm
\epsffile{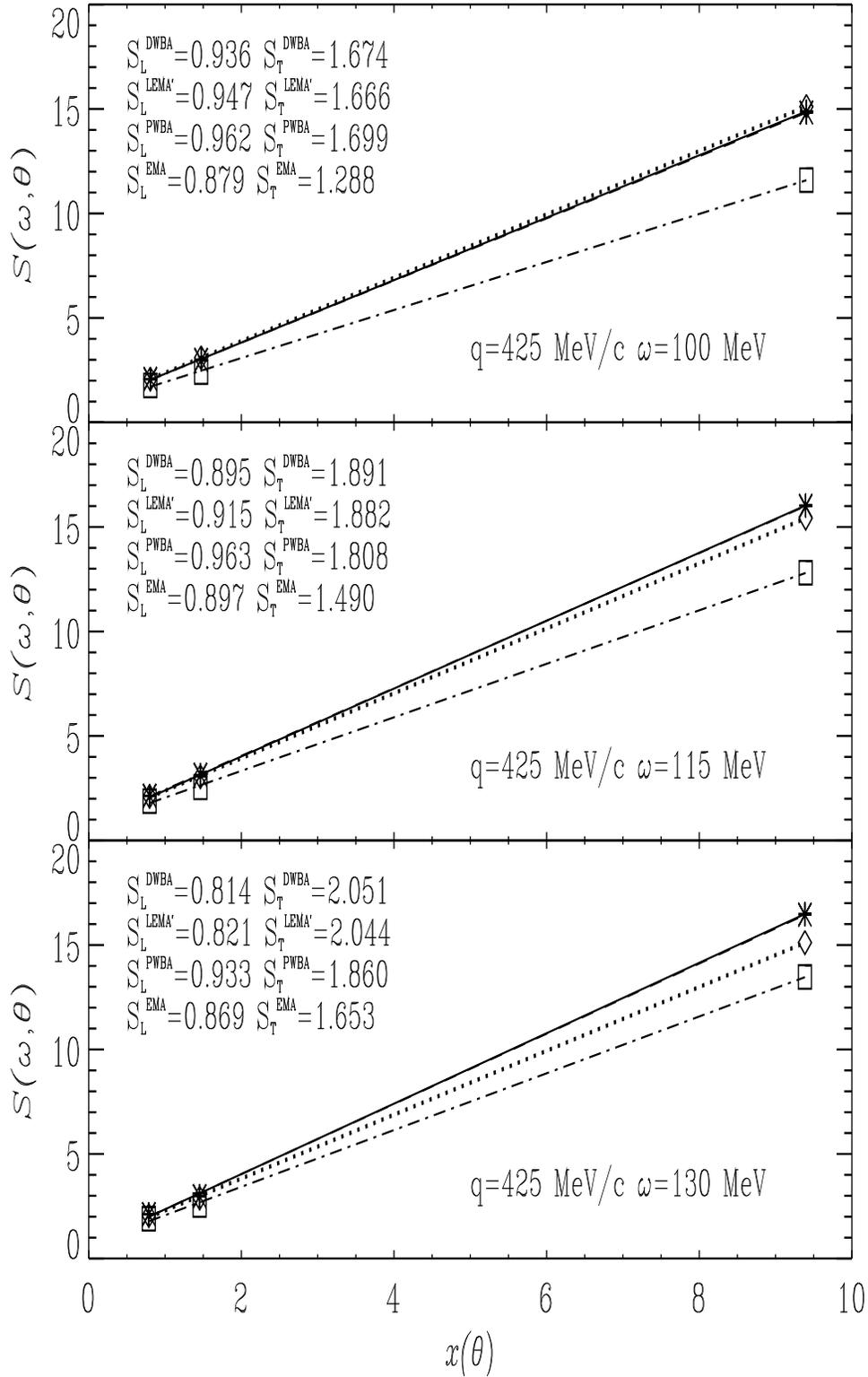}}
\noindent\hspace{5mm}\vspace{10mm}\box\figxa
\caption{Rosenbluth separation plot of the cross section for 
$^{208}Pb$ at $q=425$ MeV/c.
The solid line is the DWBA result, dashed line for the 
LEMA$^{\prime}$, the dotted line for the PWBA, and the dash-dotted 
line for the EMA. The unit of the total structure function is 
MeV$^{-1}$.}
\label{sep}
\end{figure}

In Fig.~(\ref{sep}), we compare the structure functions extracted 
using our calculations of $(e,e')$ on $^{208}Pb$ using various 
treatments of Coulomb distortion for a momentum transfer of 
$q=425$ MeV/c and three different values of energy transfer around 
the peak of the cross section.
In each case we find a very good fit to a straight line.
Furthermore, we see that the intercept and slope extracted using the 
full DWBA and LEMA$^{\prime}$ calculations agree within $2\%$ in 
all cases.
The EMA and plane wave (PWBA) results clearly are in disagreement 
with the DWBA results.

Given this result there are several ways to analyse experimental 
$(e,e')$ data.
If one has a model for the process under investigation, one can 
calculate the $r$-dependent Fourier transforms of Eqs.~(\ref{lemal'}) 
and ~(\ref{lemat'}) and compare to  the experimentally measured 
cross section.
Or, one could make a Rosenbluth separation to obtain the 
LEMA$^{\prime}$ structure functions as a function of energy 
transfer $\omega$ and asymptotic momentum transfer $q$.
Although these structure functions clearly have some dependence on 
electron energy and scattering angle in addition to their 
dependence on $q$ and $\omega$, in our model of the quasielastic 
process this dependence is not very strong and we recommend that 
in comparing a theoretical model with the extracted LEMA$^{\prime}$ 
structure functions that electron energies and angles in the same 
range as used in the experiment be used.

Lacking a model for the process being measured, one could take a  
simple model with suitable geometry and calculate the ratio of the 
cross section calculated with PWBA to one calculated with 
LEMA$^{\prime}$ and renormalize the measured cross section data by 
multiplying by this ratio.
The resulting pseudo PWBA ``data'' could then be separated by a 
Rosenbluth procedure and would produce structure functions in terms 
of the Fourier transforms of the current components.
Clearly this procedure introduces some error if the model amount of 
longitudinal and transverse contributions are not close to the 
amounts in the process being measured since LEMA$^{\prime}$ treats 
them somewhat differently.
However, this difference is not too large  and seems to be the best 
one can do.

Finally one could assume that the theoretical model has the correct 
kinematics and spatial dependence, but that the magnitude of the 
longitudinal and/or transverse portions of the model is not correct.
That is, one could use the model to calculate the longitudinal and 
transverse contributions to the cross section and then multiply each 
by a scale factor to be determined by making a least squares fit to 
the cross section data.

\section{Comparison With Experimental Data }

\begin{figure}[p]
\newbox\figha
\setbox\figha=\hbox{
\epsfysize=130mm
\epsfxsize=130mm
\epsffile{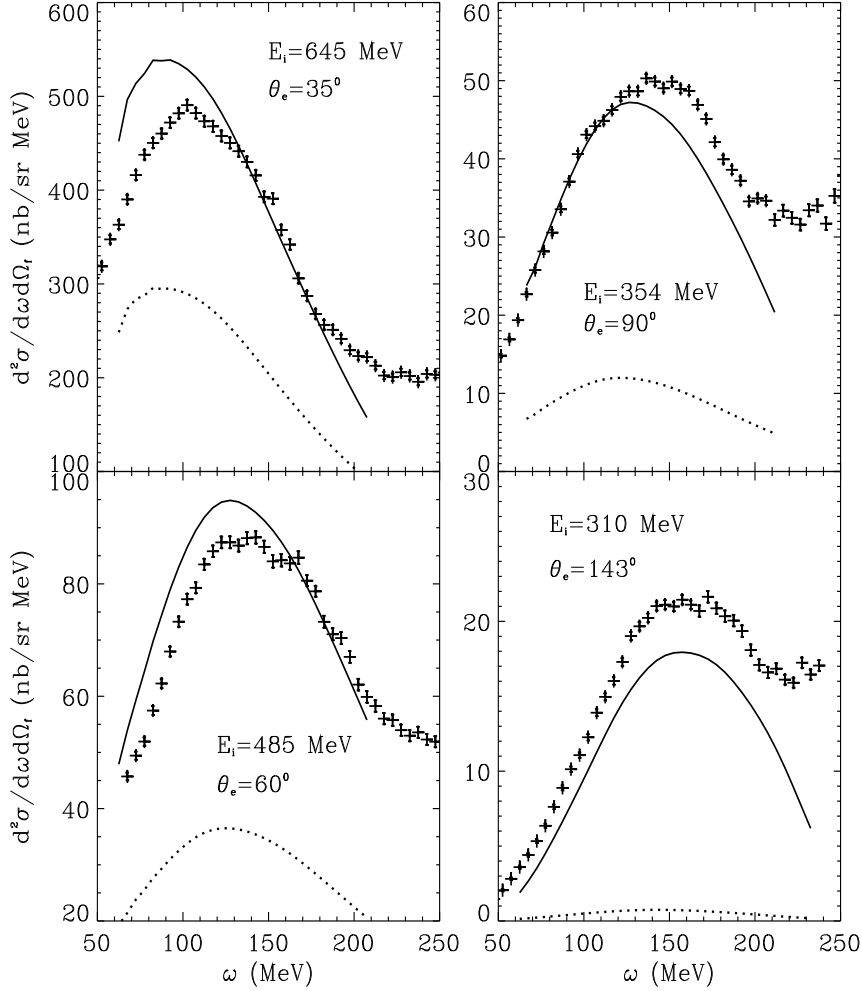}}
\noindent\hspace{20mm}\vspace{10mm}\box\figha
\caption{The differential cross section for ${^{208}}Pb(e,e')$. 
The solid line is the LEMA$^{\prime}$ result and the dotted line 
shows the longitudinal contribution. The data are from Ref.[8]}
\label{clesa}
\end{figure}

In Fig~(\ref{clesa}), we compare our theoretical results based on 
the relativistic ``single particle'' model~\cite{jin92} calculated 
with LEMA$^{\prime}$ to the Saclay data~\cite{saclay} for several 
electron angles (${\theta}_{e}=143^{0}, 90^{0}$, $60^{0}$ and 
$35^{0}$).
The solid line is our model result for the cross section while the 
dotted line shows the longitudinal contribution to the cross section.
Clearly the longitudinal contribution is quite small except for the 
forward electron scattering angle of $35^{0}$ where it represents 
slightly over $50\%$ of the cross section.
Pion production is not included in our model, so the behavior at 
large $\omega$ is not expected to agree with the data.
Clearly the agreement between the data and the calculation is not 
very good, which can be contrasted with quite good agreement between 
our calculations and the $(e,e')$ data from Bates on 
$^{40}Ca$~\cite{yates}.
Further, these may be a suggestion of longitudinal suppression if 
coupled with a transverse enhancement. 

\begin{figure}[p]
\newbox\figha
\setbox\figha=\hbox{
\epsfysize=130mm
\epsfxsize=130mm
\epsffile{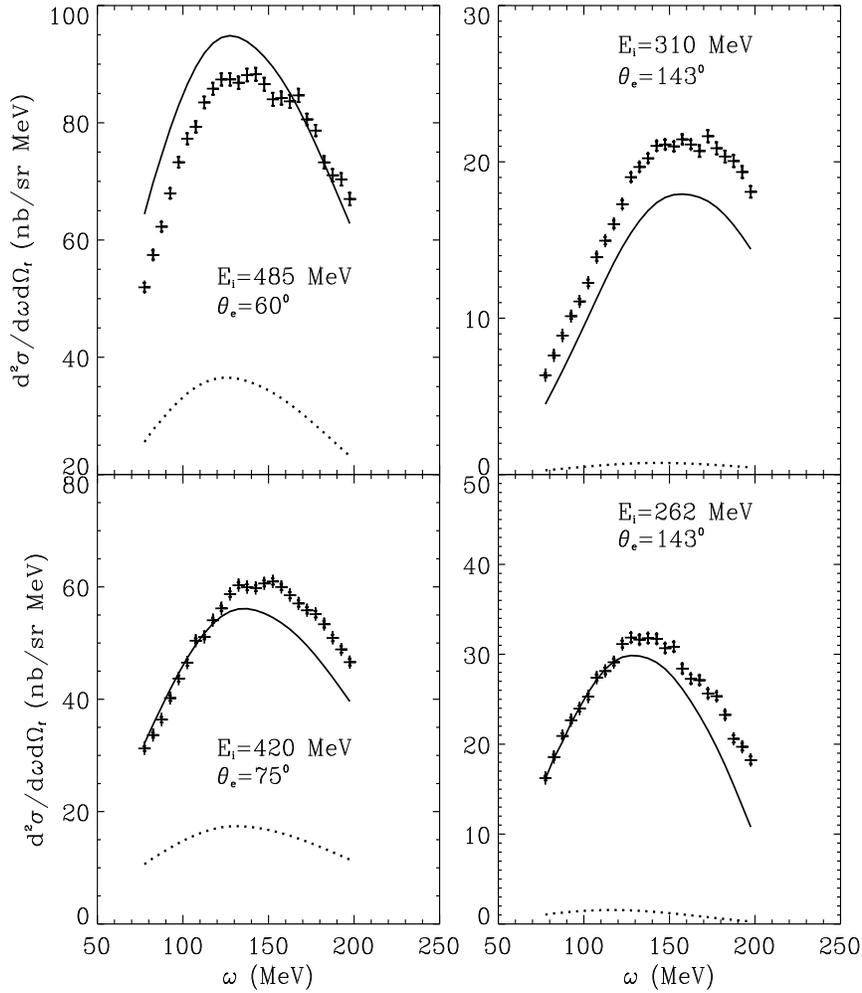}}
\noindent\hspace{20mm}\vspace{10mm}\box\figha
\caption{The inconsistent comparison with LEMA$^{\prime}$ and
Saclay data.}
\label{fitcros}
\end{figure}

We selected all of the Saclay data with energy transfer between 100 
MeV and 200 MeV which had a clearly defined quasielastic peak in the 
cross section(218 data points), and used our model with 
LEMA$^{\prime}$ to calculate the longitudinal and transverse 
contributions to the cross section for each kinematical point in 
the data set.
We performed a linear least squares fit by scale factors in front of 
the longitudinal and transverse contributions.
This fit produced a factor in front of the longitudinal term of 0.69 
and in front of the transverse term of 1.25.
However, the fit is not very good since the ${\chi}^{2}$ per data 
point is 60.  The only conclusion we can make is that we do not find 
a $50\%$ suppression of the longitudinal contribution in $^{208}Pb$.
Furthermore, it appears to us that the experimental data do not 
scale quite correctly.
In Fig.~(\ref{fitcros}) we show four sets of experimental data along 
with our model.
Two of the sets are at forward angles with quite similar kinematics 
and two are at backward angles with similar kinematics.
One of the forward angle data sets falls significantly below our 
model while the other is above.
It would be difficult to modify the longitudinal and transverse 
strength in our model to get such a dramatic shift in behavior.
For the larger angle case, the contribution of the longitudinal 
terms are negligible and again the kinematics do not vary so much 
between the two cases.
The 310 MeV result lies far above our calculation while the 
262 MeV result is only slightly above our calculation. 

\section{Conclusion}

We have developed a simple way of including Coulomb distortion 
in the M\"oller potential for inelastic electron scattering 
from medium and heavy nuclei.
In this paper we have applied a simplified version of this 
approximation to   $(e,e')$ reactions from medium and heavy nuclei 
in the quasielastic region.   
The previously used Effective Momentum Approximation (EMA) disagrees 
with the distorted wave analysis  for nuclei as heavy as $^{208}Pb$, 
while the Local Effective Momentum Approximation with an enhancement factor and an $ad-hoc$ 
correction to the longitudinal term (LEMA$^{\prime}$) reproduces the 
full DWBA calculation very well.
Our most important finding is that LEMA$^{\prime}$ allows the cross 
section to be separated into a longitudinal and a transverse 
contribution.
However,  the resulting structure functions depend on $r$-dependent 
Fourier transforms of the transition current components.
We recommend several different procedures for using LEMA$^{\prime}$ 
for the analysis of experimental data including one where distortion 
effects can be applied to the experimental data and ``plane-wave'' 
structure functions can then be extracted from the corrected data.
 
We analysed the quasielastic $(e,e')$  data from Saclay  on 
$^{208}Pb$ using a relativistic single particle model and 
LEMA$^{\prime}$.
We do not find agreement with the data and even if we vary the 
longitudinal and transverse contributions in our model do not agree 
with the data.
This is to be contrasted with our excellent agreement~\cite{yates} 
with quasielastic data on $^{40}Ca$ and rough agreement with 
quasielastic data~\cite{blat86} on $^{238}U$.
We strongly recommend  that additional  $(e,e')$ experiments on 
medium and heavy nuclei in the quasielastic region be carried out.
The treatment of Coulomb corrections is no longer a serious 
hindrance to the analysis of such experiments.

\begin{center}
ACKNOWLEDGMENTS
\end{center}

We thank the Ohio Supercomputer Center in Columbus for many hours
of Cray Y-MP time to develop this calculation and to perform the
necessary calculations. This work was supported in part by the U.S.
Department of Energy under Grant No. FG02-87ER40370.

\end{document}